\documentclass[twoside,journey]{IEEEtran}
\usepackage{makecell}
\usepackage{hyperref}
\usepackage{array}
\usepackage{graphicx,amssymb,amsmath}
\usepackage{multicol}
\usepackage[noadjust]{cite}
\usepackage{subfigure}
\usepackage{graphicx}
\usepackage{float}
\usepackage{url}
\usepackage{stfloats}
\usepackage{amsthm,pifont}
\usepackage{flushend}
\usepackage{cases,subeqnarray}
\usepackage{bm,multirow,bigstrut}
\usepackage{amsmath, amsthm, amssymb}
\usepackage{textcomp}
\usepackage{latexsym,bm}
\usepackage{booktabs}
\usepackage{xcolor}
\usepackage{mathtools}
\usepackage{dsfont}
\usepackage{extarrows}
\usepackage{epsfig}
\usepackage{epsfig}
\usepackage{epstopdf}
\usepackage[noend]{algpseudocode}
\usepackage{algorithmicx,algorithm}
\usepackage{svg}
\theoremstyle{plain}

\theoremstyle{plain}

\usepackage{amsmath}

\usepackage{tablefootnote}
\usepackage{algorithmicx}

\newcommand{\circled}[1]{\normalsize{\textcircled{\scriptsize{#1}}}\normalsize\;}
\IEEEoverridecommandlockouts

\DeclareUnicodeCharacter{3000}{~}

\begin{document}
\title{Blockchain-Empowered Lifecycle Management for AI-Generated Content (AIGC) Products in Edge Networks}

\author{Yinqiu~Liu,
      Hongyang~Du,
      Dusit~Niyato,~\IEEEmembership{Fellow,~IEEE},
      Jiawen~Kang,
      Zehui~Xiong,
      Chunyan~Miao,~\IEEEmembership{Fellow,~IEEE},
        Xuemin (Sherman)~Shen,~\IEEEmembership{Fellow,~IEEE},
        and Abbas Jamalipour,~\IEEEmembership{Fellow,~IEEE}% <-this % stops a space

  \thanks{Y. Liu, H. Du, D. Niyato, and C. Miao are with the School of Computer Science and Engineering, Nanyang Technological University, Singapore (e-mail: yinqiu001@e.ntu.edu.sg, hongyang001@e.ntu.edu.sg, dniyato@ntu.edu.sg, and ascymiao@ntu.edu.sg).}% <-this % stops a space
  \thanks{J. Kang is with the School of Automation, Guangdong University of Technology, China (e-mail: kavinkang@gdut.edu.cn)}
  \thanks{Z. Xiong is with the Pillar of Information Systems Technology and Design, Singapore University of Technology and Design, Singapore (e-mail: zehui xiong@sutd.edu.sg)}
  \thanks{X. Shen is with the Department of Electrical and Computer Engineering, University of Waterloo, Canada (e-mail: sshen@uwaterloo.ca)}
  \thanks{A. Jamalipour is with the School of Electrical and Information Engineering, University of Sydney, Australia (e-mail: a.jamalipour@ieee.org)}
  }
\maketitle

\begin{abstract}
The rapid development of Artificial Intelligence-Generated Content (AIGC) has brought daunting challenges regarding service latency, security, and trustworthiness.
Recently, researchers presented the edge AIGC paradigm, effectively optimize the service latency by distributing AIGC services to edge devices.
However, AIGC products are still unprotected and vulnerable to tampering and plagiarization.
Moreover, as a kind of online non-fungible digital property, the free circulation of AIGC products is hindered by the lack of trustworthiness in open networks. 
In this article, for the first time, we present a blockchain-empowered framework to manage the lifecycle of edge AIGC products.
Specifically, leveraging fraud proof, we first propose a protocol to protect the ownership and copyright of AIGC, called Proof-of-AIGC.
Then, we design an incentive mechanism to guarantee the legitimate and timely executions of the funds-AIGC ownership exchanges among anonymous users.
Furthermore, we build a multi-weight subjective logic-based reputation scheme, with which AIGC producers can determine which edge service provider is trustworthy and reliable to handle their services.
Through numerical results, the superiority of the proposed approach is demonstrated.
Last but not least, we discuss important open directions for further research.
\end{abstract}


% \fi
\IEEEpeerreviewmaketitle

\section{Introduction}
As an emerging technique, Artificial Intelligence-Generated Content (AIGC) has attracted significant attention from both academia and industry \cite{hongyang}. 
Instead of manually generating the content, AIGC enables the automatic creation (e.g., writing an essay, composing a song, and drawing a picture) using machine learning techniques such as Generative Adversarial Networks (GAN) and diffusion models. 
Consequently, we can acquire massive high-quality multimodal content while significantly saving on the required labor. 
Since 2014, AIGC has experienced rapid development and has been widely adopted in 3D gaming, voice assistants, video processing, etc. \cite{AIGCapplication}.

However, the current centralized AIGC framework suffers from high service latency.
For instance, to generate an image on \textit{Hugging Face} platform (\textit{https://huggingface.co/spaces}) using \textit{Stable Diffusion} model, users have to wait for 40–60 seconds.
The reasons are twofold.
Firstly, AIGC inference is complicated and time-consuming.
In the above example, the Stable Diffusion model creates images from scratch by conducting denoising operations gradually, which takes around 20–30 seconds.
Moreover, the queueing latency is also considerable (20–30 seconds in our example) since massive service requests congest one central server.

Recently, researchers have presented the idea of edge AIGC, which deploys AIGC generation services on edge devices \cite{hongyang}.
By distributing services to numerous edge devices which are close to users, service latency can be effectively reduced.
Meanwhile, the robustness gets increased due to the elimination of single-point-failure.
Moreover, users can customize AIGC services, e.g., sharing their background, locations, or characters with edge devices to generate personalized content accordingly.
Finally, since the users directly communicate with edge devices, personal information can be protected from leakage.
Although enjoying these advantages, the following challenges exist in deploying edge AIGC.
\begin{itemize}
    \item As digital property on the Internet, AIGC products are vulnerable to tampering and plagiarization (the tampering and plagiarization are shown in Section III).
    \item The economic system of AIGC is complicated. Without a mechanism guaranteeing that all the participants can benefit from AIGC circulation and obtain their deserved revenue legitimately, the generation, distribution, and trading of AIGC products will be discouraged.
    \item Recall that the generation services become distributed in edge AIGC. Therefore, as Edge Service Providers (ESPs) show significant heterogeneity in terms of model configuration and service quality, the users can hardly select reliable ESPs for their tasks.
\end{itemize}

Fortunately, blockchain provides available solutions for these issues.
As a distributed ledger, blockchain can construct trustworthiness among anonymous participants by maintaining an immutable and traceable history \cite{Lightchain}.
Moreover, smart contracts make blockchain programmable, enabling the on-chain deployment of arbitrarily complex mechanisms (e.g., two-phase locks and incentive mechanisms).
Consequently, the status and trading of AIGC products can be monitored on-chain, eliminating the security and trustworthiness problems.
In 2022, Oben AI published the proposal of \textit{AIGC chain} (\textit{https://www.aigcchain.io/about}), which allows users to contribute resources for training distributed AIGC models and acquiring rewards.
As the first blockchain for AIGC, however, this project is still under development and far from completing the whole ecosystem.
Moreover, it only uses blockchain as a crowdsourcing platform for generating AIGC, while the distribution and trading of AIGC are unprotected.

In this article, we propose the blockchain-empowered AIGC product lifecycle management in edge networks.
Specifically, we first define ``\textbf{\textit{AIGC product lifecycle}}" and discuss four major concerns regarding lifecycle management.
To help AIGC products defend malicious attacks, a Proof-of-AIGC mechanism is proposed, using fraud proofs to deal with plagiarization.
Given the complex economic system of AIGC, we further equip our framework with an on-chain incentive mechanism based on Hash Time Look (HTL) \cite{lock}.
With guaranteed and timely revenue issuance, the circulation of AIGC can be motivated and incentivized.
Finally, noticing the heterogeneity of ESPs, we enable AIGC producers to select the ESPs based on their accumulated reputation, which is modeled by Multi-weight Subjective Logic (MWSL) method \cite{reputation}.
\textit{To the best of our knowledge, this is the first work discussing the issues and solutions of AIGC product lifecycle management.}
Our contributions are summarized as follows:
\begin{itemize}
    \item We present Proof-of-AIGC mechanism. Different from Proof-of-X (e.g., Proof-of-Semantics \cite{yijingmag}), a challenge scheme is implemented, thus deregistering plagiarized AIGC products and protecting users' copyright.
    \item We propose an incentive mechanism with one-way incentives and two-way guarantees. The former encourages users to participate in managing the AIGC product lifecycle, and the latter ensures the atomic executions of AIGC trading, i.e., fund-ownership exchanges.
    \item We design a reputation-based ESP selection strategy. By calculating and sharing reputation, users can easily quantify the trustworthiness of numerous heterogeneous ESPs and assign their tasks to the most reliable one.
\end{itemize}

\section{AIGC: Current Progress, Lifecycle Management, and Concerns}
In this section, we first review the development of AIGC.
Then, we show the AIGC product lifecycle in edge networks.
Finally, important security and circulation concerns existing in the AIGC product lifecycle are discussed.

\subsection{Development of AIGC}
AIGC is an emerging generation diagram after Professional-Generated Content and User-Generated Content.
As the name suggests, the development of AIGC is driven by progresses in AI research.
Before 2010, machines can hardly generate high-quality content due to the limited capability of deep learning models. 
Since 2014, various generative neural networks have been presented, such as GAN and variational autoencoders.
Consequently, AIGC enters a period of rapid development.
In 2020, OpenAI published the \textit{Generative Pre-trained Transformer-3} (GPT-3) model, supporting multiple text generation tasks, e.g., mechanism translation and report creation \cite{GPT3}. 
Two years later, the diffusion-based \textit{DALL-E-2} model is presented.
Based on the text description given by users, DALL-E-2 can generate high-quality realistic images automatically.
Apart from text-to-text and text-to-image generation, AIGC is widely adopted in video processing, gaming, voice assistants, etc.
Moreover, it is regarded as a building block for many revolutionary techniques, including Web3, metaverse, digital twin, even the future 7G \cite{AIGCMeta}.
\begin{figure*}[htpb]
\centering
\includegraphics[width=17.3cm, height=5.3cm]{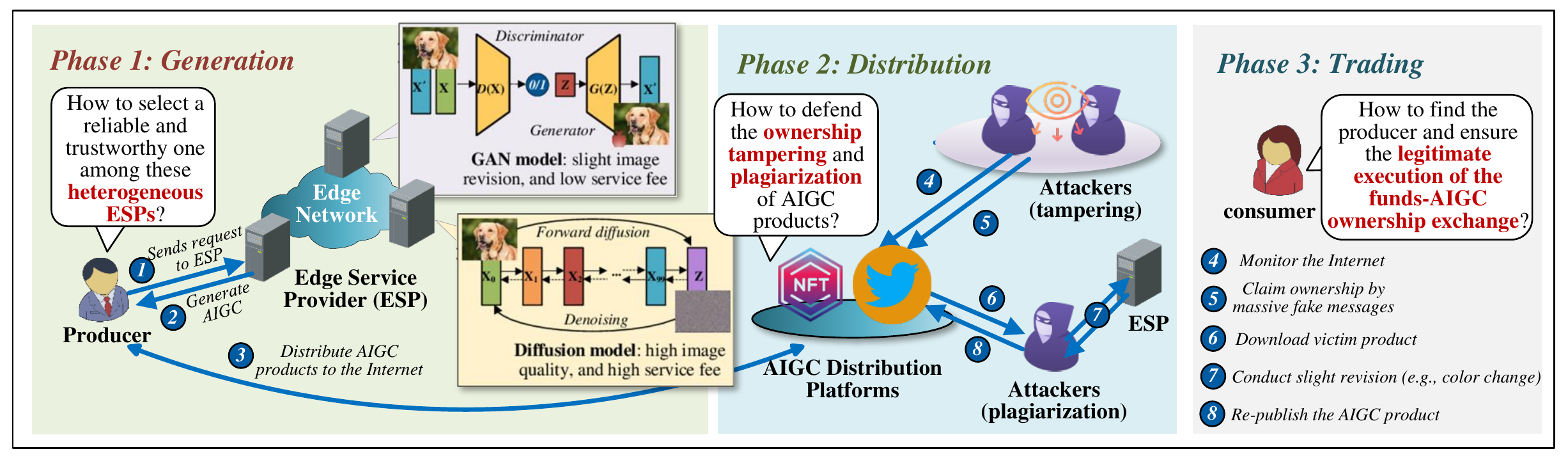}
\vspace{-0.18cm}
\caption{The AIGC product lifecycle and its important concerns.}
\label{ledger architecture}
\end{figure*}

\subsection{AIGC Product Lifecycle Management in Edge Networks}
Traditionally, AIGC models are operated by centralized servers, such as \textit{Hugging Face} platform.
In this case, massive users send requests to the central server, wait in line, and receive the services.
Researchers attempt to deploy AIGC services in edge networks to avoid request congestion and optimize service latency.
Compared with central servers, edge devices also have enough computing resources to conduct AIGC inference and are closer to users.
Therefore, the users can communicate with devices with lower transmission latency.
Moreover, since AIGC services are distributed to multiple edge devices, the waiting latency can be significantly decreased.
Nonetheless, the current research only covers the generation of AIGC products.
As a kind of non-fungible online property like NFT \cite{NFT}, each AIGC product has its ownership, copyright, and value.
Accordingly, the protection and management of AIGC products should cover their whole lifecycle.
Next, we define the concept of ``AIGC product lifecycle".

The entire AIGC product lifecycle has three phases, namely generation, distribution, and trading (see Steps \circled{1}-\circled{3} in Fig. 1).
Taking text-to-image generation as an example, the primary process of each phase is described below.
\begin{itemize}
    \item \textbf{Generation:} Producers, with insufficient physical resources, pack prompts, i.e., interesting and accurate text descriptions, and requirements in a request and sends them to ESPs (Step \circled{1}). Edge devices serve as ESPs, providing AIGC generation services for clients using local well-trained AIGC models (Step \circled{2}). Since AIGC generation is time-consuming and takes computing resources, ESPs can claim fees from producers.
    \item \textbf{Distribution:} After generation, the producers acquire the ownership of the AIGC products. Consequently, they have the right to distribute these products to social media or AIGC platforms through edge networks (Step \circled{3}).
    \item \textbf{Trading:} Since AIGC products are regarded as a novel kind of non-fungible digital properties, they can be traded. The trading process can be modelled as a fund-ownership exchange between two parties.
\end{itemize}
During such a lifecycle, several issues are yet to be addressed. 
As shown in Fig. 1, firstly, the ownership and copyright of AIGC products are vulnerable on the Internet.
Meanwhile, the producers also encounter problems in choosing reliable ESPs.
Finally, the legitimate trading of AIGC products among anonymous participants is unsolved.
In the following part, we discuss these concerns in detail.

%Although AIGC has shown great potential to be deployed in edge networks, some challenges are yet to be addressed.
%In this section, we discuss four major concerns regarding the security and circulation of edge AIGC.

\subsection{Security Concerns}
Since AIGC products are published on open networks, various kinds of attacks threaten them \cite{ownership}.
Here, we illustrate two crucial attacks targeting the AIGC products, namely the tampering of ownership and the plagiarization of AIGC.
Note that other attacks, such as denial-of-service and injection, can also destroy AIGC \cite{injection}.
Nevertheless, since they are general-purpose attacks and have been well-elaborated, we do not cover them in this article.

\subsubsection{Tampering of Ownership}
Taking text-to-image AIGC as an example, first, Steps \circled{1}-\circled{3} in Fig. 1 illustrate its lifecycle.
For conducting ownership tampering, the attackers generally deploy many robots to monitor closely the Internet and find high-quality AIGC products timely (Step \circled{4}).
After selecting the victim image, the attacker, assisted by its robots, distributes massive messages to re-publish the image, pretending that the image is its original (Step \circled{5}).
Since the attacker can broadcast information more rapidly, the consumers have a high probability of first reading the information offered by the attacker.
If so, the ownership of the victim image can be regarded as successfully tampered.

\subsubsection{Plagiarization of AIGC}
Compared with ownership tampering, the plagiarization of AIGC is harder to be detected.
In this case, the attacker will not directly claim ownership of the victim image.
Instead, it downloads the high-quality victim image, conducts some slight revision (e.g., adding noise or changing the colors of some objects), and publishes it as a brand new AIGC product (Step \circled{6}-\circled{8}).
Since such revision is much easier and cheaper than generating AIGC images from scratch, the attacker can make significant profits.
Moreover, it can even repeat this strategy, i.e., using one original image to generate a series of duplicates with little difference, thus further increasing its gains.

\subsection{Circulation Concerns}
Apart from security concerns, to realize the free circulation of AIGC, we also encounter two challenges.

\subsubsection{Heterogeneity of ESPs}
The lifecycle of every AIGC product starts from generation, i.e., using well-trained AIGC models to create content based on producers' requirements.
Nonetheless, ESPs in edge networks show great heterogeneity in model and service quality.
Taking Fig. 1 as an example, one ESP is equipped with \textit{Stable Diffusion} \cite{Diffusion}, the state-of-the-art AIGC model.
The training of Stable Diffusion is called forward diffusion, i.e., smoothly perturbing the original image data by adding noise.
The corresponding training time exceeds 150,000 hours on 256 Nvidia A100 GPUs, at a cost of US\$600,000.
In contrast, another ESP in Fig. 1 only has a simple GAN model.
The quality of the resulting content generated by these two ESPs is significantly different.
However, since ESPs may lie to producers, they cannot determine which ESPs are trustworthy.
%Furthermore, even with advanced AIGC models, ESPs may only provide semifinished products to the producers or maliciously delay the services to maximize profits.

\subsubsection{Issuance of the Deserved Revenue}
Nowadays, we are experiencing the evolution from Web2 to Web3.
In the Web3 era, everyone owns the content he/she generates.
Correspondingly, all the contributions to maintain, distribute, and enrich the community should be rewarded.
Nevertheless, ensuring that all the deserved revenue can be issued timely is challenging, especially in the AIGC scenario, whose economic system is complex.
Given the high costs of AIGC generation, the computing power and time invested by ESPs should be rewarded with fees.
Meanwhile, the producers are only willing to pay if they are guaranteed to receive the AIGC products on time.
Likewise, AIGC trading also involves a two-way guarantee of whether the producer and consumer can obtain the funds and AIGC ownership, respectively.
However, on the public Internet, two parties of transactions can hardly build trustworthiness.
Such a concern might discourage the producers from distributing and trading products, thus blocking the free circulation of AIGC products.

From the above discussion, we can observe that the difficulty of AIGC lifecycle management originates from two issues, i.e., i) the intrinsic venerability of AIGC as a kind of digital non-fungible property and ii) the lack of trustworthiness on the Internet. Fortunately, as an immutable ledger and trust maker, blockchain can effectively solve these two issues.
\begin{figure*}[htpb]
\centering
\includegraphics[width=0.93\textwidth]{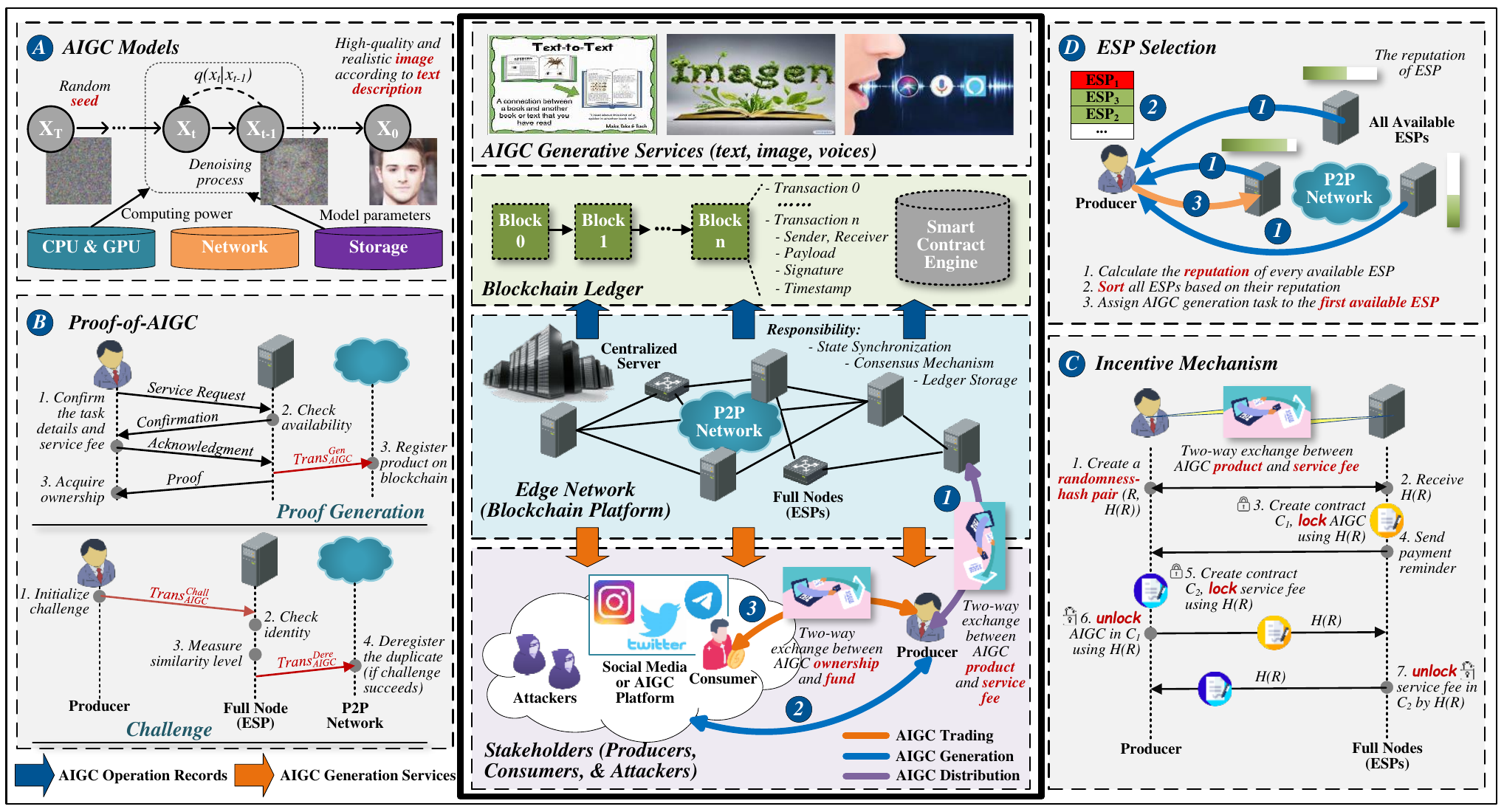}
\caption{The blockchain-empowered framework for AIGC product lifecycle management. Part A represents the AIGC models operated by ESPs. Parts B, C, and D illustrate the Proof-of-AIGC (demonstrated in Section III-B), incentive mechanism (demonstrated in Section III-C), and reputation-based ESP selection (demonstrated in Section IV), respectively.}
\label{ledger architecture}
\end{figure*}

\section{Blockchain-Empowered AIGC Lifecycle Management}
\subsection{Framework Overview}
The proposed blockchain-based framework for AIGC product lifecycle management is shown in Fig. 2. In the following part, we introduce this framework in terms of stakeholders, blockchain platform, and on-chain mechanisms.

\subsubsection{Stakeholders}
The entire AIGC product lifecycle in edge networks involves four types of stakeholders in total, namely producers, ESPs, consumers, and attackers.
\begin{itemize}
    \item \textbf{Producer:} Producers initialize the lifecycle of an AIGC product. Due to resource limitations, they only propose prompts (e.g., interesting and accurate text descriptions in text-to-image AIGC) and then request for ESPs to complete the generation tasks. After the generation, they become the first owners of the resulting products and have the right to publish and sell them.
    \item \textbf{ESP:} ESPs (e.g., edge servers) have enough resources to save well-trained AIGC models and generate content (see Fig. 2, Part A). Therefore, they can provide content generation services for producers. However, given the complexity of AIGC generation, ESPs can charge producers based on the time and computing power that they invest to the tasks.
    \item \textbf{Consumer:} After distribution, the AIGC product will be viewed by numerous people, some of whom may buy it. Such viewers are called consumers. During the lifecycle of an AIGC product, it might experience multiple times of trading with different consumers.
    \item \textbf{Attacker:} Attackers can launch ownership tampering and AIGC plagiarization to disturb the normal operations of AIGC products and make profits.
\end{itemize}

\subsubsection{Blockchain Platform}
In our framework, blockchain has two major functions: i) providing a traceable and immutable ledger and ii) supporting on-chain mechanisms.
To this end, every phase of the AIGC product lifecycle will be recorded by transactions, whose basic format is \textit{Trans\;(Sender, Receiver, Payload, Timestamp, Signature)}.
Note that the payload is different depending on the specific types of events.
Transactions are packed into blocks and submitted to the blockchain network, a distributed Peer-to-Peer (P2P) network.
The participants of the P2P network, named full nodes, conduct a consensus mechanism for block verification.
Finally, valid blocks can be appended to the ledger and saved by all full nodes in parallel.
Since everyone preserves a ledger copy, the attackers have to revise at least 50\% copies for tampering history, which is almost impossible.
In addition, we can easily trace any historical events by traversing the ledger.
Moreover, to support complex on-chain mechanisms, a turing-complete smart contract engine is deployed.

Among all participants, ESPs serve as full nodes and are responsible for message synchronization, block verification, and ledger storage.
Given the resource limitation, producers and consumers act as clients, relying on ESPs to access the blockchain services.
Note that the consensus mechanism in our blockchain is delegated Proof-of-Stake \cite{DPoS}, in which ESPs deposit stakes and take turns to create blocks.
In this case, the attackers need to manipulate 50\% ESPs for launching 51\% attacks.
Moreover, the deposited stakes will be locked if their malicious attacks are detected.

\subsubsection{On-chain Mechanism}
The framework is equipped with three on-chain mechanisms for different purposes.
Firstly, we design the Proof-of-AIGC mechanism to defend plagiarization (see Fig. 2, Part B).
To protect the funds-AIGC ownership exchange, we further implement an incentive mechanism based on HTL (see Fig. 2, Part C).
Finally, we present the reputation-based ESP selection, which effectively schedules AIGC generation tasks among ESPs (see Fig. 2, Part D).

\subsection{Proof of AIGC}
\vspace{-0.03cm}
As shown in Fig. 2, Part B, the Proof-of-AIGC consists of two phases, namely proof generation and challenge.

\subsubsection{Proof Generation}
Proof generation intends to register AIGC products on blockchain.
We still take text-to-image AIGC as an example.
For generating an image, the producer first sends a corresponding request to an ESP (ESP selection strategy is discussed in Section IV).
The request format is \textit{(Text description, service fee, expected time)}.
After receiving the service request, the ESP checks its availability and decides whether to accept the task.
If the expected time and service fee are acceptable, it conducts a handshake with the producer (see Fig. 2, Part B).
Then, the image creation can be conducted by the ESP, using well-trained AIGC models.

After generating the image, ESP initializes a transaction $Trans_{AIGC}^{Gen}$\textit{(Sender, Receiver, Payload, Timestamp, Signature)}.
The Payload format is \textit{(Product index, Metadata, Challenge expiration)}, in which \textit{Product index} is calculated by hash function and is regarded as the unique identity for the AIGC product. 
\textit{Metadata} contains the basic information of the AIGC product.
Such a transaction will go through the verification and be recorded by the blockchain.
Finally, the ESP will send the image to the producer, with a copy of $Trans_{AIGC}^{Gen}$.
$Trans_{AIGC}^{Gen}$ can be regarded as a proof, which not only registers the AIGC product, but also claims its ownership by setting \textit{Receiver} as producer's address.
Given the immutability of blockchain ledger, the concerns about ownership tampering can be effectively addressed.
%Since producers only publish the registered images, even though attackers can distribute massive fake messages, they cannot tamper the on-chain registration records preserved in all ESPs.
%However, the risk of AIGC plagiarization still exists.
Next, we demonstrate the challenge mechanism to help producers defend the AIGC plagiarization.

\subsubsection{Challenge}
Proof-of-AIGC follows the principle of fraud proof.
In other words, our blockchain assumes that all AIGC products are original work in the proof generation phase.
However, the information recorded in $Trans_{AIGC}^{Gen}$ enables producers to challenge any on-chain AIGC product that they believe copies their own work.
If the challenge succeeds, the duplicate will be deregistered, thus protecting the copyright of the real producer. 
Next, we illustrate the challenge workflow. 

Suppose that the producer has created and published an AIGC product (called original product).
Then, it surfs the Internet and finds an AIGC product which is significantly similar to its own work (called duplicate).
In this case, it can initialize the challenge process by sending a transaction $Trans_{AIGC}^{Chall}$ with the payload \textit{($\textit{Product}_1$, Product $\textit{index}_1$, $\textit{Product}_2$, Product $\textit{index}_2$, Pledge deposit)}.
Here, $\textit{Product}_1$ ($\textit{Product}_2$) and \textit{Product $\textit{index}_1$} (\textit{Product $\textit{index}_2$}) represent the content and indexes of the original product (duplicate), respectively.
We consider that the duplicates will also be registered on blockchain because consumers will only buy the AIGC products with clear proof.
After receiving $Trans_{AIGC}^{Chall}$, the ESPs will conduct the following four steps:
\begin{itemize}
    \item \textbf{Step 1:} \textbf{Fetch the proofs}. The $Trans_{AIGC}^{Gen}$ of both the original product and the duplicate will be fetched from local ledger. Recall that the format of $Trans_{AIGC}^{Gen}$ is \textit{(Sender, Receiver, Payload, Timestamp, Signature)}.  
    \item \textbf{Step 2:} \textbf{Check the identity of the challenger.} The ESPs verify challenger's signature in $Trans_{AIGC}^{Chall}$ using \textit{Receiver} public key in $Trans_{AIGC}^{Gen}$. If signature verification is successful, it can prove that the challenger is indeed the owner of the original product.
    \item \textbf{Step 3:} \textbf{Measure the similarity between the original product and the duplicate.} Firstly, ESPs conduct hash operations on $\textit{Product}_1$ and $\textit{Product}_2$ and check whether the hashes match \textit{Product $\textit{index}_1$} and \textit{Product $\textit{index}_2$}, respectively. If so, they conduct the similarity measurement using three well-established metrics, namely image histogram, perceptual hash, and difference hash. Note that the metrics can be changed for other AIGC scenarios.
    \item \textbf{Step 4:} \textbf{Check the results}. If the similarity level exceeds the threshold in any two metrics, the challenge can be regarded as successful. Otherwise, the challenge fails. 
\end{itemize}

If the challenge succeeds, the ESPs create and send a transaction $Trans_{AIGC}^{Dere}$ with the payload \textit{(Product $\textit{index}_2$, Pledge deposit, Similarity)}, where \textit{Similarity} is defined as a three-element tuple \textit{(histogram, phash, dhash)}. 
$Trans_{AIGC}^{Dere}$ aims to deregister the duplicate by pointing out its product index. 
Moreover, it unlocks the pledge deposit provided by the challenger. 
Recall that the challenge is initialized by $Trans_{AIGC}^{Chall}$, whose \textit{Sender} is obviously the challenger's address.
However, the \textit{Receiver} address of $Trans_{AIGC}^{Chall}$ does not belong to any participant.
Instead, it is a special system account for locking the pledge deposit provided by challenger.
The motivation for requiring pledge deposit is to restrict challengers from launching challenges arbitrarily, since the challenge process causes extra burden to the blockchain.
Nonetheless, such deposit can be waived if the challenge happens before the pre-defined \textit{Challenge expiration} in $Trans_{AIGC}^{Gen}$.
For example, if the original product is registered in the 5\textit{th} block and \textit{Challenge expiration} is 20, the challenge will be free from the 6\textit{th} to the 20\textit{th} block. 
From the 21\textit{st} block, the challenger can only withdraw its deposit if it successfully proves that a duplicate is mistakenly registered on blockchain.
Otherwise, the locked pledge deposit will be regarded as a service fee and be used to reward the next block creator.

\subsection{Incentive Mechanism}
The economic system of AIGC is complicated because it accommodates different stakeholders, which conduct transactions with each other frequently.
Thus, we should guarantee that: i) all the stakeholders can be incentivized to manage the AIGC lifecycle; ii) the funds-AIGC ownership exchanges can be conducted legitimately without repudiation.
To this end, an on-chain incentive mechanism is presented.

\subsubsection{One-way Incentives}
One-way incentives are automatically issued to the ESPs which maintain the ledger and provide blockchain services.
Recall that our blockchain adopts delegated Proof-of-Stake as the consensus mechanism, where ESPs take turns to generate new blocks.
During each round of block generation, the generator can include a coinbase transaction to reward itself.
The \textit{Sender} and \textit{Receiver} addresses of such coinbase transactions are the system account and generator's public key address, respectively.
For the specific reward value, it can be set according to the target system inflation rate. 
Note that there is no transaction fee in our incentive mechanism.
Hence, block generator just packs pending transactions by the first-come-first-serve strategy.

\subsubsection{Two-way Guarantee}
As mentioned before, during both AIGC generation and trading, there exist two-way exchanges between fund and ownership.
However, people might hesitate to conduct such exchanges, since they cannot guarantee that the other party will strictly follow its promise.
To build mutual trust and facilitate AIGC circulation, we design a two-way guarantee protocol using HTL (Hash Time Lock) as a part of our incentive mechanism.

Take the two-way exchange happening in AIGC generation phase as an example.
In this case, the ESP grants the producer the ownership of its AIGC product, and the producer pays the pre-configured service fee.
To do so, we implement a smart contract with two atomic operations named \texttt{lock} and \texttt{release}.
As shown in Fig. 2, Part C, during the handshake process described in Section IV-B, the producer creates a randomness $R$ and sends its hash $H(R)$ to ESP.
When $Trans_{AIGC}^{Gen}$ is recorded on the blockchain, a corresponding contract instance $C_1$ will be created by ESP immediately.
$C_1$ calls \texttt{lock} function to lock the ownership storing in $Trans_{AIGC}^{Gen}$ using $H(R)$.
Only the one with $R$ can release the lock.
Meanwhile, the ESP sends a payment reminder to the producer.
Receiving the bill, the producer sends a payment transaction with the payload \textit{(Balance)}, where \textit{Balance} should be equal to the pre-configured service fee.
Then, it also initializes its own contract instance $C_2$, which locks the fund in the payment transaction by $H(R)$.
Up till now, both fund and AIGC ownership are on-chain.

Then, the secure exchange between fund and AIGC ownership can be conducted.
Firstly, the producer unlocks the AIGC ownership by calling \texttt{release} operation of $C_1$, with an input $R$.
$C_1$ will check whether the hash of $R$ matches $H(R)$ and unlock the $Trans_{AIGC}^{Gen}$ if $H(R)$ is correct.
Since such a process exposes $R$ to $C_1$, the owner of $C_1$, i.e., ESP, can also release the fund locked by $C_2$ using $R$.
To prevent participants from intentional delay, we further add an expiration to the smart contract.
Consequently, if they fail to unlock the properties on time, the lock will become permanent and the corresponding transactions will be discarded.
Clearly, such a protocol guarantees the atomic and timely executions of the exchange process.

\subsection{Security Analysis}
Recall that in Section II, we point out four concerns of the AIGC product lifecycle, namely ownership tampering, AIGC plagiarization, non-guaranteed exchanges, and ESP heterogeneity.
Firstly, our Proof-of-AIGC registers every AIGC products on chain.
In this case, even though attackers can distribute massive fake messages to ``claim" their ownership, they can hardly launch 51\% attacks (as mentioned in Section III-A) or tamper the registration history preserved in all full nodes.
Additionally, the challenge scheme provides the standard procedure for producers to defend AIGC plagiarization and retrieve their copyright.
Furthermore, the incentive mechanism guarantees that all the funds-AIGC ownership exchanges can be conducted strictly following the pre-confirmed contracts.
Next, we address the final concern, i.e., ESP heterogeneity, by presenting a reputation-based ESP selection.

\section{Reputation-Based ESP Selection}
\subsection{Problem Statement}
Recall that AIGC services are distributed to numerous edge devices in our framework.
Hence, each producer can access multiple heterogeneous ESPs simultaneously.
In this case, selecting a reliable ESP for the specific task becomes a problem.
Traditionally, producers can select the most familiar ESP, i.e., the one with which they have traded the most times, to minimize the potential risk.
However, such strategies may lead to an imbalanced workload among ESPs, thus increasing the service latency on busy ESPs.
Meanwhile, the computing resources of idle ESPs will be wasted.

To solve this problem, we implement a reputation-based ESP selection scheme in our framework.
Specifically, it sorts all available ESPs according to their reputation, which is calculated by Multi-weight Subjective Logic (MWSL) \cite{reputation}.
We intend to achieve three goals: i) helping producers select the most reliable ESP for each AIGC generation task; ii) balancing the workload among multiple ESPs, thereby reducing the overall service latency; iii) encouraging ESPs to complete the assigned tasks timely and honestly, since a negative reputation will directly affect their profits.
\begin{figure}[tpb]
\centering
\includegraphics[width=9.5cm, height=6.8cm]{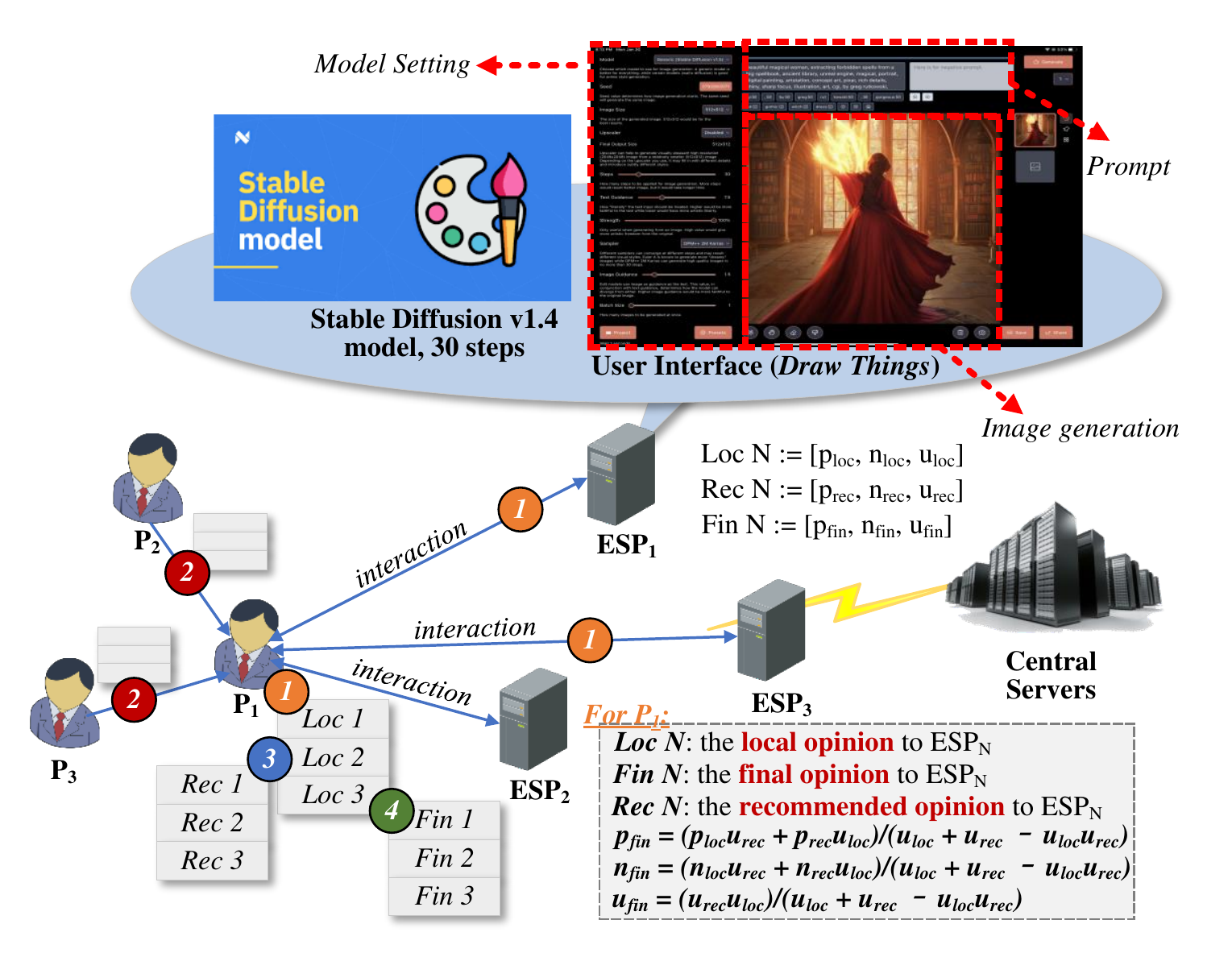}
\caption{The reputation calculation process (from the perspective of producer $P_1$) and the illustration of AIGC services.}
\label{ledger architecture}
\end{figure}

\subsection{Reputation Based on Multi-weight Subjective Logic}
As shown in Fig. 2, Part D, producers select ESPs by the following steps: i) calculate the reputation of all available ESPs, ii) sort candidate ESPs according to their latest reputation, and iii) assign the AIGC generation task to the ESP with the highest reputation. 
Note that the item $ESP_1$ is marked red because it denies the service request. 
In this case, the producer traverses the reputation table and re-sends the request to the next candidate, i.e., $ESP_3$.
Next, we demonstrate the reputation calculation based on MWSL. 

As shown in Fig. 3, MWSL utilizes the term ``opinion" to denote the basic items for reputation calculation.
Suppose that our edge AIGC has three producers ($P_1$-$P_3$) and three ESPs ($ESP_1$-$ESP_3$).
Firstly, for a given producer, say $P_1$, if it has direct interactions with these ESPs, $P_1$'s evaluation of them is called local opinions.
Meanwhile, considering that $P_2$ and $P_3$ may also have the experience for interacting with these ESPs, their evaluation should also be taken into account.
From the perspective of $P_1$, the evaluation of $ESP_1$-$ESP_3$ from $P_2$ and $P_3$ are called recommended opinions.
Here, an interaction refers to the entire process from sending service request, to confirming AIGC generation order, and to acquire AIGC products.
The opinion is defined as a three-element vector [$p, n, u$], where $p$ and $n$ represent the proportion of \textit{positive} and \textit{negative} interactions in all interaction attempts, respectively. 
$u$ (from 0 to 1) indicates the uncertainty level between producer and ESP.
According to MWSL, $u$ is set manually according to the communication quality.

Although recommended opinions make reputation calculation more comprehensive, the hidden subjectivity might affect the fairness.
For instance, if $P_2$ once suffered an unexpected high latency from $ESP_1$, it may regard all subsequent interactions as negative.
To mitigate the effect caused by subjectivity, for each producer, say $P_1$, an overall opinion averaging all the received recommended opinions will be generated.
Moreover, since $P_2$ and $P_3$ have different familiarity degrees with ESPs, the weight of their recommended opinions is also different.
The detailed reputation calculation process is:
\begin{itemize}
    \item \textbf{Step 1:} \textbf{Generate local opinions}. Every producer updates its local opinion for every ESP (see Step \circled{1} in Fig. 3). 
    \item \textbf{Step 2:} \textbf{Synchronize information.} Producers share the latest local opinions. Assisted by blockchain, they can pack their opinions into transactions for secure sharing.
    \item \textbf{Step 3:} \textbf{Calculate overall opinion} Each producer collects all received recommended opinions and averages them as the overall opinion. Note that the opinions are weighted before calculating the average. For any recommended opinion from $P_n$ to $ESP_n$, the weight is $\alpha_1 \times Familiarity + \alpha_2 \times Value$. $Familiarity$ is defined as the number of historical interactions between $P_n$ and $ESP_n$. $Value$ equals the total service fee for these interactions. Finally, $\alpha_1$ and $\alpha_2$ are two weighting factors satisfying $\alpha_1$ + $\alpha_2$ = 1. Notably, the more interactions have been conducted, the larger the weight.
    \item \textbf{Step 4:} \textbf{Calculate reputation.} Every producer combines its local opinion with overall opinion and achieves the final opinion [$p_{fin}, n_{fin}, u_{fin}$]. The corresponding equation is shown in Fig. 3. Finally, reputation is measured by $p_{fin}$ + $u_{fin}$ $\times$ $n_{fin}$.
\end{itemize}

After reputation calculation, producers take Steps \circled{2}-\circled{3} in Fig. 2, Part D, and select an ESP. 
Clearly, our reputation scheme successfully achieves all the design goals. 
Firstly, it quantifies the trustworthiness of ESPs. 
Hence, producers can easily determine which ESP is more reliable. 
In addition, producers do not need to only rely on the most familiar ESP, thereby alleviating the potential service congestion. 
Finally, since the reputation records are store on-chain and clear to all participants, ESPs are encouraged to provide high quality AIGC services for maximizing their profits.

\begin{figure}[tpb]
\centering
\includegraphics[width=0.66\columnwidth]{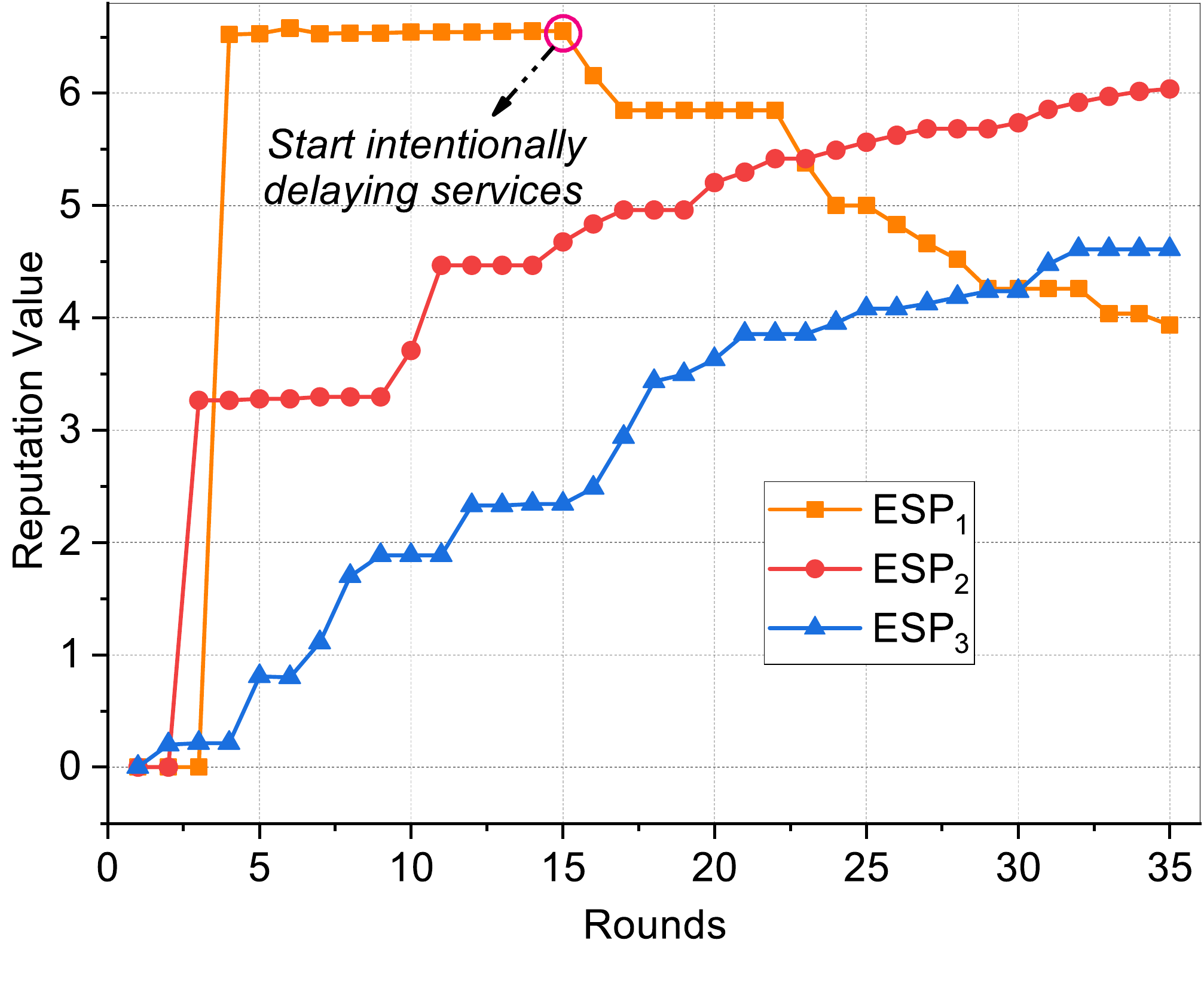}
\vspace{-0.4cm}
\caption{The reputation trends of three ESPs (from the perspective of a random producer).}
\vspace{-0.4cm}
\end{figure}

\begin{figure}[tpb]
\centering
\includegraphics[width=0.65\columnwidth]{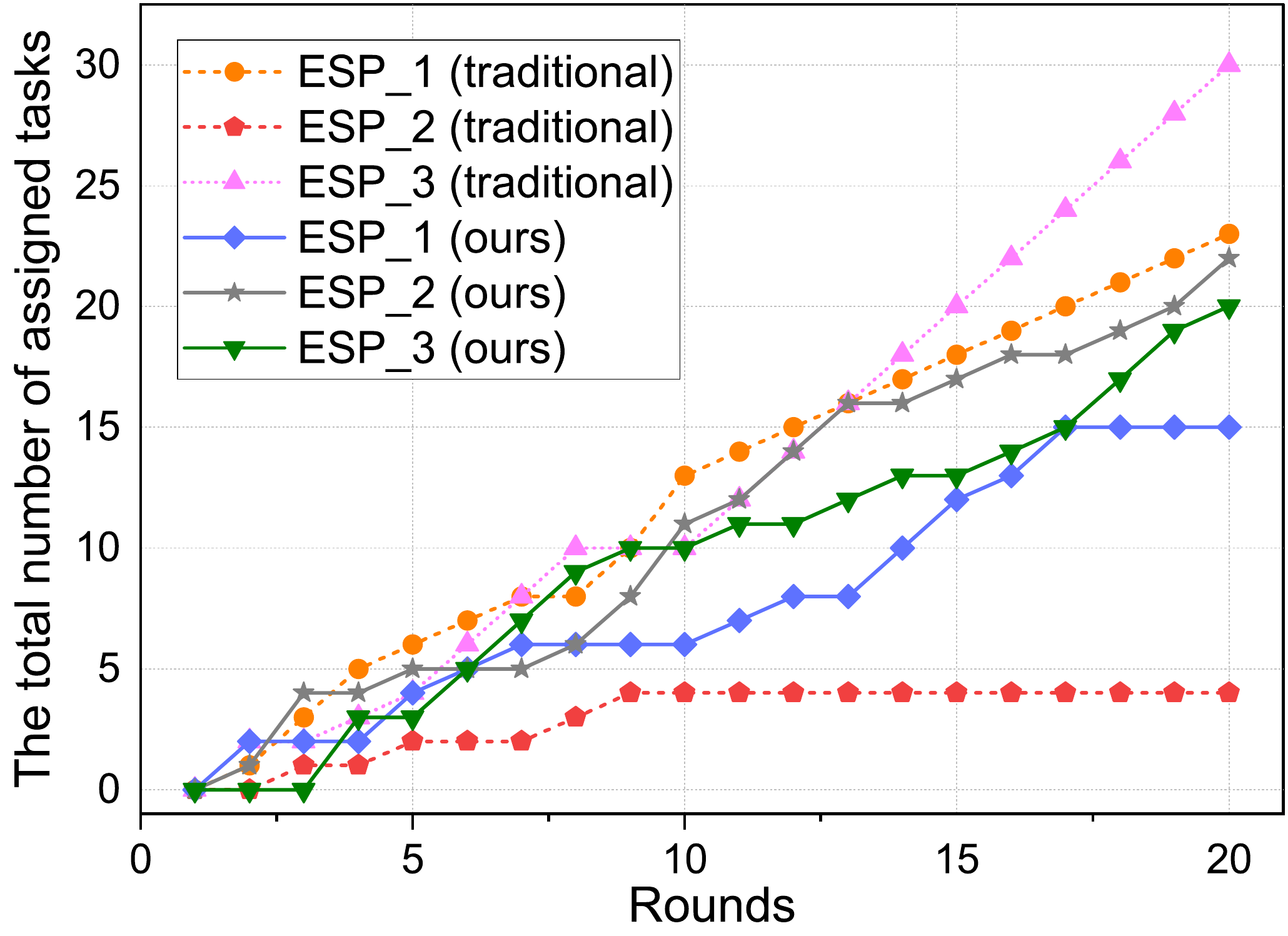}
\caption{The total number of assigned tasks of three ESPs.}
\vspace{-0.6cm}
\end{figure}

\vspace{-0.2cm}
\subsection{Numerical Results}
To prove the validity of the proposed methods, we implement a demo of our AIGC lifecycle management framework and deploy the reputation-based ESP selection on it\footnote{https://github.com/Lancelot1998/AIGCLifecycleManagement}.
As shown in Fig. 3, the testbed consists of three ESPs (served by three virtual machines on Apple MacBook Pro with 8-Core Intel Core i9 CPU and AMD Radeon Pro 5500M GPU) and three producers (served by iPhones).
The AIGC services are supported by \textit{Draw Things} application (\textit{https://drawthings.ai/}).
Factors $\alpha_1$ and $\alpha_2$ are set as 0.35 and 0.65, respectively.
Additionally, $u_{loc}$ is fixed to 0.5 \cite{reputation}.
For each producer, it marks one interaction as ``negative" if ESP fails to return the AIGC proof within the pre-confirmed time.
The service quality (i.e., the probability of receiving positive opinions) of $ESP_1$, $ESP_2$, and $ESP_3$ are 95\%, 70\%, and 55\%, respectively.
Finally, after acquiring the ESPs' reputation, producers can utilize \textit{Softmax} function to determine the probability for selecting each ESP.

Firstly, Fig. 4 illustrates the reputation trends of three ESPs.
During the 1\textit{st}-14\textit{th} rounds, all ESPs accumulate reputation.
Given the high service quality, the reputation of $ESP_1$ directly reaches to the top and stays stable, while $ESP_2$ and $ESP_3$ gradually increase their reputation by providing more positive interactions.
From the 15\textit{th} round, we let $ESP_1$ intentionally delay the AIGC services.
Corresponding, its reputation drops dramatically, since more negative interactions are reported.
In contrast, since $ESP_2$ and $ESP_3$ acquire the chance to handle more tasks, their reputation keeps increasing.
We conclude that the proposed reputation scheme can effectively quantify the trustworthiness of ESPs.
In this way, the producers can easily judge which ESP is the most reliable.
On the other hand, ESPs are also supervised to keep performing honestly.

Then, Fig. 5 shows ESPs' workload under different ESP selection methods.
Here, we suppose that all three producers request for AIGC services with the same frequency, and we let them randomly select ESPs during the first 5 rounds.
From the 6\textit{th} round, two ESP selection methods are tested, namely traditional method and the proposed reputation-based method.
Recall that traditionally, the producers tend to assign tasks to their most familiar ESPs.
As a result, the workload among ESPs is imbalanced, causing long service latency.
As shown in Fig. 5, most AIGC generation tasks congest in $ESP_3$, while the computing power of $ESP_2$ becomes wasted.
Assisted by reputation, the producers can qualitatively evaluate the trustworthiness of ESPs and no longer need to rely on their empirical judgement.
Consequently, the workload among ESPs is effectively balanced.
Note that since there is no similar work regarding blockchain-empowered AIGC in the literature, we do not set a baseline to compare with.

\vspace{-0.1cm}
\section{Future Direction}
\subsection{Blockchain-Based AIGC Governance}
The rapid development of AIGC greatly enriches the Internet content, but it also brings \textit{deepfake} \cite{deepfake}.
Deepfake refers to synthetic media in which a person in an existing image or video is replaced with someone else's likeness.
According to thesentinel (\textit{https://thesentinel.ai/}), the number of deepfake videos online has jumped from 14,678 in 2019 to 145,277 in 2021.
Moreover, leveraging advanced AIGC models, such as GAN and autoencoders, deepfake is becoming more and more realistic and harder to be identified.
Given the security property of blockchain, it can help AIGC against deepfake.
For example, some distributed governance organization can be deployed on-chain, thus conducting the AIGC supervision and deepfake identification.
However, since identifying deepfake requires off-chain knowledge, how to effectively bridge blockchain and physical AIGC is worth exploring.

\vspace{-0.1cm}
\subsection{Distributed AIGC Model Training}
This article mainly focuses on the AIGC product lifecycle.
The AIGC model construction lifecycle, including model training, fine-tuning, and inference, is also a meaningful research topic.
For instance, since the training of diffusion models is time-consuming and resource-intensive, new algorithms and frameworks for building the distributed AIGC model training are worth studying.
In this way, the computing power in the entire edge network can be exploited and thus significantly improve the training speed.
Meanwhile, blockchain can be applied to protect the security of the training process and reward the users who contribute their resources fairly.

\vspace{-0.1cm}
\subsection{Metaverse}
AIGC is a building block for metaverse, since it can create numerous multimodal content for rendering immersive and realistic virtual worlds \cite{metaverse}.
For example, the text-to-3D AIGC allows machines to collect the background, locations, and characters of users, thereby generating personalized avatars in the metaverse environment.
Although such a process brings high QoE and immersiveness, some sensitive personal information might be leaked.
Since blockchain has shown great strength in protecting data storage and sharing, the metaverse-oriented AIGC storage, access control, and sharing based on blockchain technique are also worth investigating.

\vspace{-0.1cm}
\section{Conclusion}
In this article, we first review the progress of AIGC and its deployment in edge networks.
Then, we point out four major concerns of the AIGC product lifecycle.
Hence, we present a blockchain-empowered framework, realizing the lifecycle management for AIGC products.
Specifically, Proof-of-AIGC solves the ownership tampering and plagiarization of AIGC products.
Additionally, an incentive mechanism is proposed to encourage the AIGC circulation.
Moreover, we design a reputation scheme to help producers select reliable ESPs, with numerical results to prove its validity.
Last but not least, we discuss future directions regarding the combination of blockchain and AIGC.

\ifCLASSOPTIONcaptionsoff
  \newpage
\fi

\bibliographystyle{IEEEtran}
\bibliography{aigcblockchain}
\vfill

\end{document}